\documentclass[conference]{IEEEtran}
\IEEEoverridecommandlockouts

\usepackage[nolist]{acronym}
\usepackage{cite}
\usepackage{amsmath,amssymb,amsfonts}
\usepackage{algorithmic}
\usepackage{graphicx}
\usepackage{textcomp}
\usepackage{hyperref}
\usepackage{cleveref}
\usepackage{xcolor}
\usepackage{array}
\usepackage{subcaption}
\usepackage{balance} 

\def\BibTeX{{\rm B\kern-.05em{\sc i\kern-.025em b}\kern-.08em
    T\kern-.1667em\lower.7ex\hbox{E}\kern-.125emX}}
    
\begin{document}

\title{Injecting Conflict Situations in Autonomous Driving Simulation using CARLA
\thanks{
This work was partially supported by the Wallenberg AI, Autonomous Systems and Software Program (WASP) funded by the Knut and Alice Wallenberg Foundation and Saab AB.}
}

\author{
\IEEEauthorblockN{Tsvetomila Mihaylova}
\IEEEauthorblockA{
\textit{School of Electrical Engineering} \\
\textit{Aalto University}\\
Espoo, Finland \\
tsvetomila.mihaylova@aalto.fi}
\and
\IEEEauthorblockN{Stefan Reitmann}
\IEEEauthorblockA{\textit{Dept. of Computer Science} \\
\textit{Lund University}\\
Lund, Sweden \\
stefan.reitmann@cs.lth.se}
\and
\IEEEauthorblockN{Elin A. Topp}
\IEEEauthorblockA{\textit{Dept. of Computer Science} \\
\textit{Lund University}\\
Lund, Sweden \\
elin\_a.topp@cs.lth.se}
\and
\IEEEauthorblockN{Ville Kyrki}
\IEEEauthorblockA{
\textit{School of Electrical Engineering} \\
\textit{Aalto University}\\
Espoo, Finland \\
ville.kyrki@aalto.fi}
}

\maketitle

\begin{abstract}
Simulation of conflict situations for autonomous driving research is crucial for understanding and managing interactions between \acp{av} and human drivers. This paper presents a set of exemplary conflict scenarios in CARLA that arise in shared autonomy settings, where both \acp{av} and human drivers must navigate complex traffic environments. We explore various conflict situations, focusing on the impact of driver behavior and decision-making processes on overall traffic safety and efficiency. We build a simple extendable toolkit for situation awareness research, in which the implemented conflicts can be demonstrated.
\end{abstract}


\section{Summary}

In Level-3 automated driving \cite{SAE2021}, the \ac{av} operates on its own for most of the journey and the driver can concentrate on other tasks. In case of a conflict situation (i.e. a situation in which the autonomous system is uncertain how to proceed), the control has to be passed to the driver. In these cases, the system needs to provide explanations that would allow the human to understand the situation and act in a short time.
In order to enable research in this scenario, we build a toolkit based on the CARLA simulator \cite{Dosovitskiy17}, that allows HRI researchers to quickly simulate conflict situations and easily conduct user studies to explore how the human reacts to different explanations, which could lead to improvements in situational awareness. We implement several conflict types, chosen based on a framework for conflict simulation \cite{gold2018}.

Our contributions are:

\begin{itemize}
    \item Implementation of conflicts situations in driving scenarios in the CARLA simulator, chosen based on their urgency and needed reaction time.
    \item A simple extendable toolkit\footnote{\url{https://github.com/aalto-intelligent-robotics/aisa-toolkit}} for experiments with situation awareness in autonomous driving using the CARLA simulator.
    \item A basic setup that can showcase the implemented conflicts, which initiates a takeover request in case of uncertain lane marking detection.
\end{itemize}

\begin{figure}[ht!]
    \centering
    \includegraphics[width=.45\textwidth]{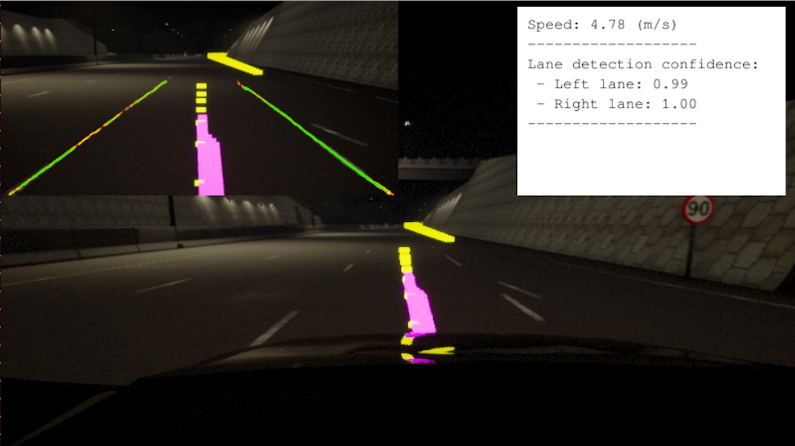}
    \caption{View from interior with detected lanes (green) and predicted path (purple).}
    \label{f:interior}
\end{figure}

\section{Background}

In this section, we discuss details of the problem of shared autonomy in the context of automated vehicles to motivate why a simulation environment for conflict situations is needed and explain our reasoning behind choosing to implement specific conflicts.

\subsection{Autonomous driving in shared autonomy}

In shared autonomy, automated systems and humans collaborate to execute tasks effectively \cite{schilling2016towards,kanda2017human}. When automated systems encounter limits defined by their design—such as environmental, locational, temporal, or traffic constraints, collectively termed \ac{odd}\cite{SAE2021}—a \ac{tor} is triggered, switching to manual control. The system must detect conflicts \cite{kuchar2000review} and provide sufficient context awareness \cite{pichler2004context, makkonen2009context, mcaree2017towards, petersen2019situational, ignatious2023analyzing} to help the human respond promptly. We focus on \ac{sae}-Level 3 (semi-autonomous driving), where the vehicle handles most of the journey while the driver can attend to other tasks \cite{SAE2021}. 
We simulate the scenario where a human is in the automated vehicle and observes the driving situation. In case of uncertainty, caused by a conflict situation, a \ac{tor} is initiated.

\subsection{Conflicts in autonomous driving}

\subsubsection{Data of conflict situation in autonomous driving}


Several works focus on simulating conflicts. Nair \textit{et al.} predict situations requiring intervention and evaluate them in the CARLA simulator using static and dynamic obstacles and adverse weather conditions~\cite{Nair2021}. Haojie \textit{et al.} simulate scenarios in the SUMO simulator, detecting conflicts based on vehicle motion~\cite{xin2024}. Reitmann \textit{et al.} develop a conflict simulation tool using Unity~\cite{AISA_HRI24}. Datasets with conflicts from crash or near-crash situations are also available~\cite{Stark2020Generation, Bashetty2020DeepCrashTest:, Nitsche2017Pre-crash}. However, to the best of our knowledge, no open-source simulation meets these criteria: enabling easy conflict simulation, allowing human observation and control (supporting situation awareness research), and offering features to enhance situation awareness.




\subsubsection{Conflict classification}
\label{ss:conflictclassification}

The most general classification of conflicts can be seen as a binary classification within the problem of reaching the limit of the \ac{odd}: first, the system reaches its technical limits, so it does not have the capabilities to find a solution for a given problem or has to break (traffic) rules. Second, the system's technical capabilities are limited due to failure, e.g. sensor malfunctions. A lot of empirical studies have been conducted to analyse single types of conflicts within this binary split. A main concern in the class of system limits is lane markings (e.g. blurred/missing lane markings, secondary lane markings because of work zones, road curvatures) \cite{melcher2015take,naujoks2014effect,zeeb2016take}. Other studies concentrated on the traffic dynamics and possible interactions with other participants, e.g. pedestrians, obstacles on the road, cut-in vehicles \cite{larsson2014learning,naujoks2015controllability,gold2013partially,doi:10.1177/1541931214581434} or environmental factors \cite{SCHOMIG20156652,MARKVOLLRATH20111134}.

A framework to build a common ground for singular conflicts in autonomous driving is proposed by Gold \textit{et al.}, where conflicts are classified according to their \textit{urgency} (time budget to solve the conflict), \textit{predictability} (dependencies to other factors), \textit{criticality} (safety risks) and \textit{driver response} (complexity of solution)~\cite{gold2018}. 
When considering conflict situations that are to be resolved by human intervention, urgency and driver response are particularly important. The chosen conflict situations implemented in our toolkit are listed in \Cref{tab:conflicts} (without values of criticality and predictability).


\begin{table}[h!]
    \centering
    \begin{tabular}{c|p{4cm}|c|c}
    No. & Name & Urgency &  Response  \\
    \hline 
    2 & Sensor failure (total) & 3 & 1-2 \\
    5 & Lane change from entrance ramp not possible & 3 & 3 \\
    7 & Road narrows (detected by on-board sensors) & 3 & 2 \\
    9 & Danger zone/obstacle ahead & 3 & 1-3 \\
    10 & Loss of reference signals (e.g. lane markings missing) & 3 & 1 \\
    \end{tabular}
    \caption{Conflicts with highest ratings for urgency and driver response from Gold \textit{\textit{et al.}} \cite{gold2018}.}
    \label{tab:conflicts}
\end{table}

\section{Purpose}

To address the problem of the lack of data for conflict situations in autonomous driving, our intention is to create a synthetic, reproducible, adaptive database covering various setups in a virtual environment. Comprehensive conflict data is vital for enhancing safety measures, ensuring regulatory compliance, understanding system limitations, advancing technology development, building public trust, and establishing comparative safety metrics in the evolving landscape of \acp{av}. We narrow down the selection of conflicts for the simulation following the classification in Gold \textit{et al.} \cite{gold2018} and choose the conflicts to implement based on higher ratings of urgency and driver response, listed in \Cref{tab:conflicts}.

The simulation of \acp{av} can be done in different ways \cite{Li_2024}, mostly based on game engines (e.g. Unity, Unreal Engine). A well-known example of a comprehensive simulator is CARLA \cite{Dosovitskiy17}, which is open-source and actively maintained, and it offers a wide range of possibilities for \ac{hmi} investigations during conflict situations in the future, e.g. with \ac{vr} integration \cite{silvera2022dreyevr}.

The different nature of the simulated conflicts requires various technical implementations in CARLA. This includes the setup and placement of the ego-vehicle on the selected map, the integration of static and dynamic actors and the definition of the environmental parameters. \Cref{tab:supported-conflicts} describes the types of implementation that are used to transfer the chosen conflicts of \Cref{tab:conflicts} to the CARLA simulation environment.


\begin{table}[h!]
    \centering
    \begin{tabular}{ m{1.5cm} | m{2.5cm} | m{3cm} }
         \textbf{Type} & \textbf{Description} & \textbf{Technical \newline Implementation} \\
         \hline
         \textbf{Conflicts location} & Spawning the ego vehicle at a specified location that we know contains a certain conflict. & Specify the map, and list of spawning points in an XML file. \\
        \hline
        \textbf{Sensor noise} & Adding noise to the ego vehicle sensors. & Implemented in the script, after the ego vehicle and its sensors are initialized. Parameter \texttt{sensor\_noise} specified in the config. \\
        \hline
        \textbf{Static obstacles} & Placing obstacles at certain or random places in the map. & Parameters in the config. \\
        \hline
        \textbf{Dynamic obstacles} & Spawning actors (vehicles and pedestrians) at certain or random places in the map. & Based on CARLA script \textit{generate\_traffic.py}, parameters for frequency and type of actors. \\
        \hline
        \textbf{Changing weather conditions} & Specify weather conditions that can disturb the vehicle sensors. & In the scenario setup, specify \texttt{WeatherId} based on the the CARLA weather presets. \\      
         \hline
    \end{tabular}
    \vspace{1em}
    \caption{Technical transfer of supported conflict types (mentioned in \Cref{tab:conflicts}) currently supported by our toolkit.}
    \label{tab:supported-conflicts}
\end{table}





\section{Characteristics}




\subsection{CARLA Baseline Setup}

The toolkit is built for CARLA $0.9.15$, which is the latest stable version of CARLA in the time of writing. The proposed Python version is $3.7.x$. 
We adapt the simulation to be able to run in first person view. 
During the simulations we display basic information about the route and the predicted path (example on \Cref{f:interior}).

\subsubsection{Controller Model}
For the control of the \ac{av} we offer the possibility for switching to manual control when the uncertainty of the model is below a configurable threshold. To demonstrate the implemented conflicts, we implement the automatic control by using a simple lane detection model \cite{aad}. This model is not state-of-the-art, but the reason why we choose it is because of its simplicity, which allows us to focus on the implementation of conflicts and quickly testing them. The toolkit is built in a way that it could be extended by implementing more complex models for automatic control.

\subsubsection{Basic Situation Awareness}
We display basic information about the driving situation, such as the vehicle speed and probability of detected lane markings; the trajectory predicted by the model is marked on the road. There is a possibility to also display the target trajectory - from the start point to a given destination (with an argument \texttt{--draw\_route}). Warning and critical messages are displayed when the lane detection is under some configurable values. When a \ac{tor} is initiated, a text message is displayed, and an audio indication can be played (specified by an argument \texttt{--audio}).

\subsection{Implemented Conflict Instances}

We rely on available CARLA environments\footnote{\url{https://carla.readthedocs.io/en/latest/core_map/}} for implementation and supplement these (where necessary) with our own user-created custom map.
We choose the conflicts to implement based on urgency and response ~\cite{gold2018}) and implement 2, 5, 7, 9 and 10, shown in \Cref{tab:conflicts}. Each of them can be implemented in several ways, and here we describe the implementations we currently support.


\subsubsection{Conflicts in CARLA's default maps}


The standard maps included in CARLA already contain a wide range of options for conflict generation. In our toolkit, the corresponding towns are loaded, target coordinates are read in and the automated vehicle is transferred to the specific conflict situation.

\paragraph{Vanishing Lane Markings}

\begin{figure}[h!]
    \centering
    \includegraphics[width=.45\textwidth]{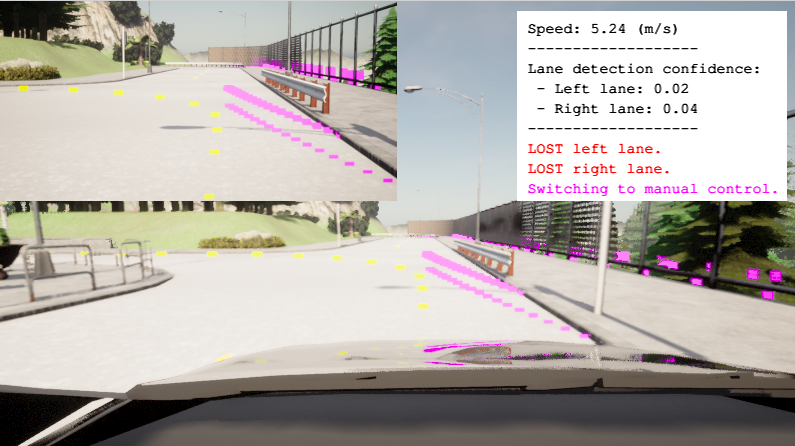}
    \caption{Vanishing lane markings in Town01. (Yellow: road indication from starting point to a given destination.)}
    \label{f:lane_markings_map}
\end{figure}

The default variant of the conflict includes changes to the quality, concealment or color of road markings. 
This conflict is present for example in CARLA's Town01 with a lead time of approx. $30$ seconds (\Cref{f:lane_markings_map}).
According to \cite{MELCHER20152867} lane markings may also change due to work zones and be replaced by secondary lane markings. We address this issue on our custom track, shown on \ref{p:vanishingcustom}.

\paragraph{Vanishing Lane Markings (Weather)}

\begin{figure}[h!]
    \centering
    \includegraphics[width=.45\textwidth]{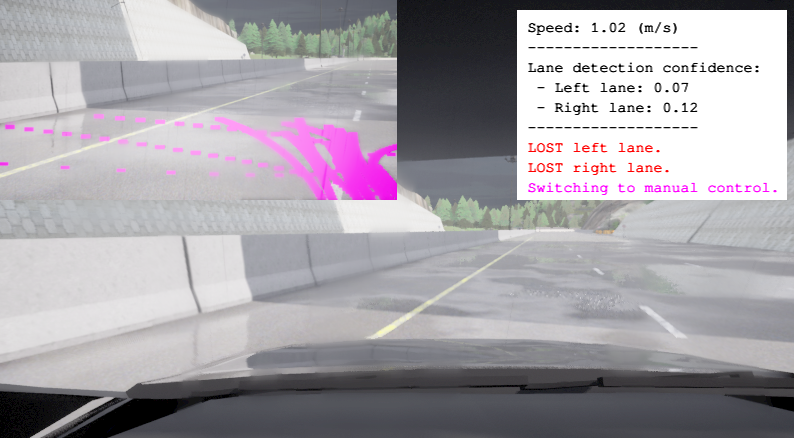}
    \caption{Vanishing lane markings under worsening weather conditions in Town04.}
    \label{f:rain}
\end{figure}

When weather conditions change, problems can occur in image-based detection even if the quality of the road markings remains the same. Weather phenomena can influence visibility, artifact formation and reflection. To illustrate this conflict, we decided on a test segment in Town04 using the \textit{HardRainSunset} weather preset. An example is shown in \Cref{f:rain}.

\paragraph{Narrowing Road}
\label{p:narrow}

\begin{figure}[h!]
    \centering
    \includegraphics[width=.45\textwidth]{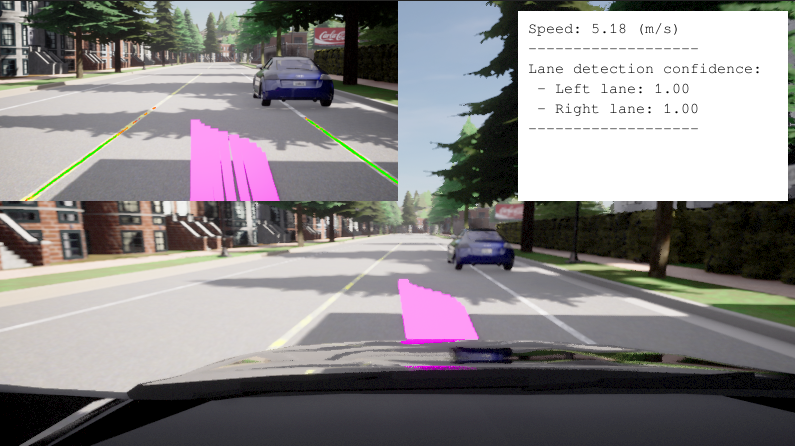}
    \caption{Camera image (on the top left of the view) with added parking vehicle, simulating a narrowing road in Town05.}
    \label{f:camera_narrow}
\end{figure}

A vehicle from CARLA's blueprint library is spawned. The model can freely be modified - also traffic signs or work zones are possible. In our example a vehicle is parked at the roadside and blocking the lane partly (example on \autoref{f:camera_narrow}). A slight evasive maneuver is needed with possibility to intersect with the opposite lane.

\paragraph{Danger Zone / Obstacle}

Similar to \autoref{p:narrow}, but the spawned vehicle is rotated and blocking the whole lane. An evasive maneuver to the opposite lane is needed. We also support dynamic obstacles, for example other cars entering our lane.


\paragraph{Sensor Failure}

\begin{figure}[h!]
    \centering
    \includegraphics[width=.45\textwidth]{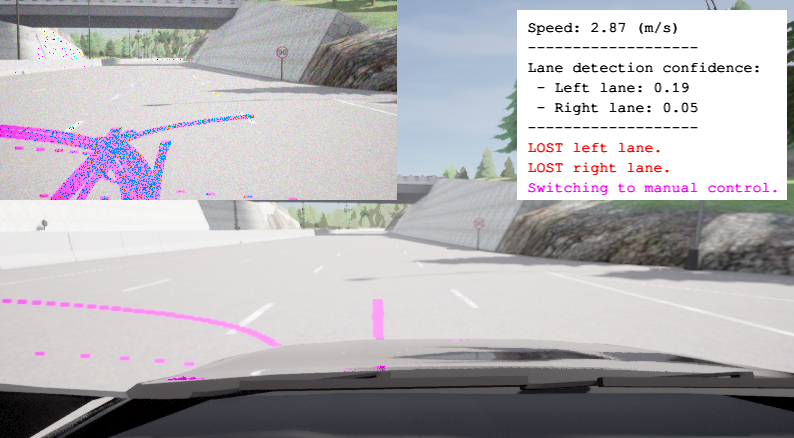}
    \caption{Camera image (on the top left of the view) with added noise, simulating sensor failure, in Town04.}
    \label{f:camera_noise}
\end{figure}

We can simulate sensor failure by adding noise to the camera (\Cref{f:camera_noise}). The amount of noise can be set in the configuration file. When the sensors are noisy, the outputs of the model are uncertain and a \ac{tor} is initiated.

\subsubsection{Conflicts in custom maps}

The maps available for CARLA offer a wide range of options for conflict simulations. However, some conflict types can only be represented with difficulty, e.g. the on-ramp conflict, as the American highway type (represented in the CARLA maps) differs from the European highway type (target platform for on-ramp situations). 

We created a custom map (closed highway loop, several inner tracks) built on the \textit{OpenDRIVE} and \textit{OpenScenario} standards containing road layouts (*.xodr) and 3D assets (*.fbx). We used the Blender \cite{blender3d} open source addon \textit{Driving Scenario Creator} \cite{BlenderDriving} proposed in \cite{RocketLoop}.




\paragraph{On-ramp Conflict}

\begin{figure}[h!]
    \centering
    \includegraphics[width=.48\textwidth]{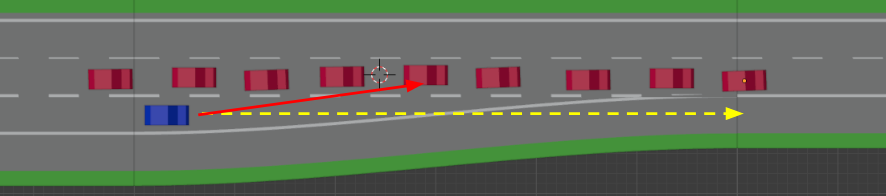}
    \caption{On-ramp conflict with intended lane-change (red) of ego vehicle (blue) and forced trajectory (yellow).}
    \label{f:onramp}
\end{figure}

\Cref{f:onramp} depicts the basic dynamic of the on-ramp conflict. If there is no space to merge into the flow of traffic, one must theoretically slow down at the end of the merging lane. 
However, this represents a considerable danger, so it is advised to use the shoulder until a switch is possible. The automated vehicle is forced to cross a solid line of the ending ramp due to the blocked lane (furthest right) of the highway, which would require a TOR or confirmation by the operator.

\paragraph{Vanishing Lane Markings (Custom Track)}
\label{p:vanishingcustom}

\begin{figure}[h!]
    \centering
    \includegraphics[width=.45\textwidth]{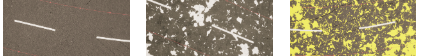}
    \caption{Decreasing lane marking quality levels in UE4.}
    \label{f:roadtext}
\end{figure}

In addition to the options available in Town01, we have added further route sections for the vanishing lane markings in the custom map (example on \autoref{f:roadtext}). These are characterized by varying quality of the marking textures (partially covered, dirty, different colors) and changing route parameters (straight, curved, winding lines). The decreasing lane marking quality leads to a decrease of the certainty of image-based models and initiates a TOR.

\section{Code \& Usage}

\paragraph{Requirements}
Our conflict simulation toolkit is based on the CARLA simulator and has been tested with version 0.9.15. In order to run our code, the simulator has to be downloaded from the CARLA repository \footnote{\url{https://github.com/carla-simulator/carla/tree/0.9.15}} and the CARLA server needs to be started (details are available in the README). The setup has been tested on Ubuntu 22.04 with an NVIDIA RTX 3080. 
Following CARLA requirements, GPU of size at least 8GB is recommended.

\paragraph{Installation and running}
The README provides instructions how to create a Conda environment with Python 3.7 and the required dependencies.
The entry point of the simulation is the script \texttt{simulation.py}, the arguments specify the conflict, controller model, and whether to use audio signal for takeover request.


\paragraph{Scenario setup}
We describe the different scenarios with an XML file. For each scenario, the following can be specified:
\begin{itemize}
    \item Map (i.e. one of the towns in CARLA), start point and destination.
    \item Weather conditions - give the id of on of the available weather presets\footnote{\url{https://carla.readthedocs.io/en/stable/carla_settings/}}.
    \item Sensor noise (the value is the $\sigma$ in adding Gaussian noise with mean $\mu=0$).
\end{itemize}


\paragraph{Loading and including a custom map}
We exported our custom map as \texttt{.blend}-file (for customization in Blender with \cite{BlenderDriving}) and \texttt{.xodr}-/\texttt{.fbx}-files for the integration in CARLA. These files can be imported following the official documentation\footnote{\url{https://carla.readthedocs.io/en/0.9.15/tuto_M_custom_map_overview/\#importation}}. The name of the imported map should match the name in the XML file (default: \textit{conflictmap}).

\paragraph{Adding new controller models}
The toolkit supports adding new controller models that inherit 
\texttt{ControllerModel}, and \texttt{ControllerModelFactory} instantiates the corresponding controller class based on an argument \texttt{--model} passed to the starting script.
The controller model accepts as input the sensor readings and outputs control. Currently, camera image and CARLA map are being sent as readings, and new sensors can be added if needed for more complex control. 


\section{Conclusion}
We present an extendable toolkit for simulating conflict situations in autonomous driving in the CARLA driving simulator. We provide implementations of several conflict types, which were chosen based on their urgency and reaction time, allowing for the handling of TOR according to recent related findings~\cite{AISA_HRI24}.
The toolkit currently has a basic interface with a simple control model, which serves the purpose of showcasing the supported conflicts. This implementation is the first step towards building an environment for further experiments regarding users'  situation awareness in autonomous driving.




\bibliographystyle{IEEEtran}
\balance
\bibliography{bib/HRI24_AISA,bib/2024_HRI_AISA_Stefan_local,bib/HRI25_AISA}

\begin{acronym}[ECU]

\acro{ann}[ANN]{Artificial Neural Network}
\acro{agtor}[AGTOR]{Attention-guiding Takeover Requests}
\acro{ai}[AI]{Artificial Intelligence}
\acro{ar}[AR]{Augmented Reality}
\acro{agv}[AGV]{Autonomous Ground Vehicle}
\acro{asv}[ASV]{Autonomous Surface Vehicle}
\acro{av}[AV]{Automated Vehicle}
\acro{cave}[CAVE]{Cave Automatic Virtual Environment}
\acro{gps}[GPS]{Global Positioning System}
\acro{hmi}[HMI]{Human Machine Interface}
\acro{lidar}[LiDAR]{Light Detection and Ranging}
\acro{ml}[ML]{Machine Learning}

\acro{odd}[ODD]{Operational Design Domain}
\acro{radar}[Radar]{Radio Detection and Ranging}
\acro{ugv}[UGV]{Unmanned Ground Vehicle}

\acro{hmd}[HMD]{Head-Mounted Display}
\acro{hmi}[HMI]{Human Machine Interface}

\acro{ndrt}[NDRT]{Non-Driving Related Task}

\acro{pid}[PID]{Proportional–integral–derivative controller}
\acro{png}[PNG]{Portable Network Graphics}

\acro{ros}[ROS]{Robot Operating System}
\acro{sa}[SA]{Situational Awareness}
\acro{sonar}[Sonar]{Sound Navigation and Ranging}

\acro{tcp}[TCP]{Transmission Control Protocol}
\acro{tor}[TOR]{Takeover Request}
\acro{tubaf}[TUBAF]{Freiberg University of Mining and Technology}
\acro{sae}[SAE]{Society of Automotive Engineers}

\acro{ue4}[UE4]{Unreal Engine 4}

\acro{vr}[VR]{Virtual Reality}
\acro{xr}[XR]{Extended Reality}
\end{acronym}

\end{document}